\documentclass[12pt]{report}
\usepackage{amsbsy,amssymb,graphicx}
\newcommand{\lam}{\mbox{$\rm \Lambda$}}% Lambda
\newcommand{\alam}{\mbox{$\rm \bar \Lambda$}} % Anti-Lambda
\begin{document}
%\begin{titlepage} 
%
\title{\hspace{7cm} DOC-2004-Jan-152\\
\vspace{6cm}
A Study of Spin-dependent Interactions with Antiprotons\\ \em
The Structure of The Nucleon \\
\vspace{15mm}
ASSIA Collaboration\\
\vspace{6cm}
}
\date{January 15, 2004}
\maketitle
%
%\newpage
%
%\begin{Authlist}
V.Abazov$^1$, G.Alexeev$^1$,  
M. Alexeev$^2$,A.~Amoroso$^2$, N.Angelov$^1$, M. Anselmino$ ^3$, S.Baginyan$^1$,
F.~Balestra$^{2}$, V.A.~Baranov$^1$, Yu.Batusov$^1$, I.Belolaptikov$^1$, 
R.~Bertini$^{2}$, N.~Bianchi$^{11}$,A.~Bianconi$^4$, R.~Birsa$^{13}$, 
T.Blokhintseva$^1$, A.Bonyushkina$^1$,
F.~Bradamante$^{13}$, A.~Bressan$^{13}$, M.P.~Bussa$^2$,
V.Butenko$^1$, M.~L.~Colantoni$^5$, M.~Corradini$^{4}$, S.~Dalla~Torre$^{13}$,
A.Demyanov$^1$, O.~Denisov$^2$, E.~De~Sanctis$^{11}$,  P.~Di~Nezza$^{11}$,
V.Drozdov$^1$, J.~Dupak$^9$, G.Erusalimtsev$^1$, L.~Fava$^5$, A.~Ferrero$^2$, L.~Ferrero$^2$,
M.~Finger$^6$, M.~Finger $^7$, V.~Frolov$^2$, R.~Garfagnini$^2$, M.~Giorgi$^{13}$,
O.~Gorchakov$^1$,
A.~Grasso$^2$, V.~Grebenyuk$^1$,  D.~Hasch$^{11}$, V.~Ivanov$^1$,
A.~Kalinin$^1$, V.A~Kalinnikov$^1$, Yu.~Kharzheev$^1$, N.V.~Khomutov$^1$, A.~Kirilov$^1$, 
Y.~Kisselev$^1$, E.~Komissarov$^1$, 
A.~Kotzinian$^2$, A.S.~Korenchenko$^1$, V.Kovalenko$^1$, N.P.~Kravchuk$^1$, N.A.~Kuchinski$^1$, 
E.~Lodi Rizzini$^{4}$, V.~Lyashenko$^1$, V.~Malyshev$^1$, A.~Maggiora$^2$, 
M.~Maggiora$^2$, A.~Martin$^{13}$, Yu.~Merekov$^1$, A.S.~Moiseenko$^1$, 
V.~Muccifora$^{11}$, A.~Olchevski$^1$, 
V.~Panyushkin$^1$, D.~Panzieri$^5$, 
G.~Piragino$^2$, G.B.~Pontecorvo$^1$, A.~Popov$^1$, S.~Porokhovoy$^1$, V.~Pryanichnikov$^1$,
M.~Radici$^{14}$, P.G.~Ratcliffe$^{12}$,
M.P.~Rekalo$^{10}$,  P.~Rossi$^{11}$, A.~Rozhdestvensky$^1$, N.~Russakovich$^1$, P.~Schiavon$^{13}$, 
O.~Shevchenko$^1$, A.~Shishkin$^1$, V.A.~Sidorkin$^1$, N.~Skachkov$^1$,
M.~Slunecka$^7$, A.~Srnka$^9$, V.~Tchalyshev$^1$, F.~Tessarotto$^{13}$,
E.~Tomasi$^8$, F.~Tosello$^2$, E.P.~Velicheva$^1$, L.~Venturelli$^4$,
L.~Vertogradov$^1$, M.~Virius$^9$ ,G.~Zosi$^2$ and N.~Zurlo$^{4}$ \\
 
%\address{
$^1$Dzhelepov Laboratory of Nuclear Problems, JINR, Dubna, Russia\\
$^2$Dipartimento di Fisica ``A. Avogadro'' and INFN - Torino, Italy  \\
$^3$Dipartimento di Fisica Teorica and INFN - Torino, Italy  \\
$^4$Universit\`{a}  and INFN, Brescia, Italy \\
$^5$Universita' del Piemonte Orientale and INFN sez. di Torino - Italy\\
$^6$Czech Technical University , Prague, Czech Republic\\
$^7$Charles University, Prague, Czech Republic\\
$^8$DAPNIA,CEN Saclay, France\\
$^9$Inst. of Scientific Instruments Academy of Sciences,Brno, Czech Republic \\
$^{10}$NSC Kharkov Physical Technical Institute, Kharkov, Ukraine \\
$^{11}$Laboratori Nazionali Frascati, INFN, Italy\\
$^{12}$Universit\`{a} dell' Insubria,Como and INFN sez. Milano, Italy\\
$^{13}$University of Trieste and INFN Trieste, Italy\\
$^{14}$INFN sez. Pavia, Italy\\
%}
%
%\end{Authlist}
%\maketitle
%
\tableofcontents
\chapter{Introduction}
\label{section:Introd}
The availability, at the new GSI facility (SIS300), of an antiproton beam with momenta larger than
$40 \, GeV/c$, that is with wave lenghts smaller than $4 \cdot 10^{-2}$ 
fm., makes possible a detailed study of the nucleon structure. 
 
A complete description of the nucleon structure requires the knowledge of parton 
distributions (PD), including also the gluon distributions, at leading twist (twist 2) and next to leading 
order (twist3). 
The definition of these functions will be recalled in (Section~\ref{section:strucfunc}).

These distributions can be obtained through the measurement of spin dependent cross-sections in a variety 
of hard processes.

\begin{figure}
\centering
\includegraphics[totalheight=8.5cm]{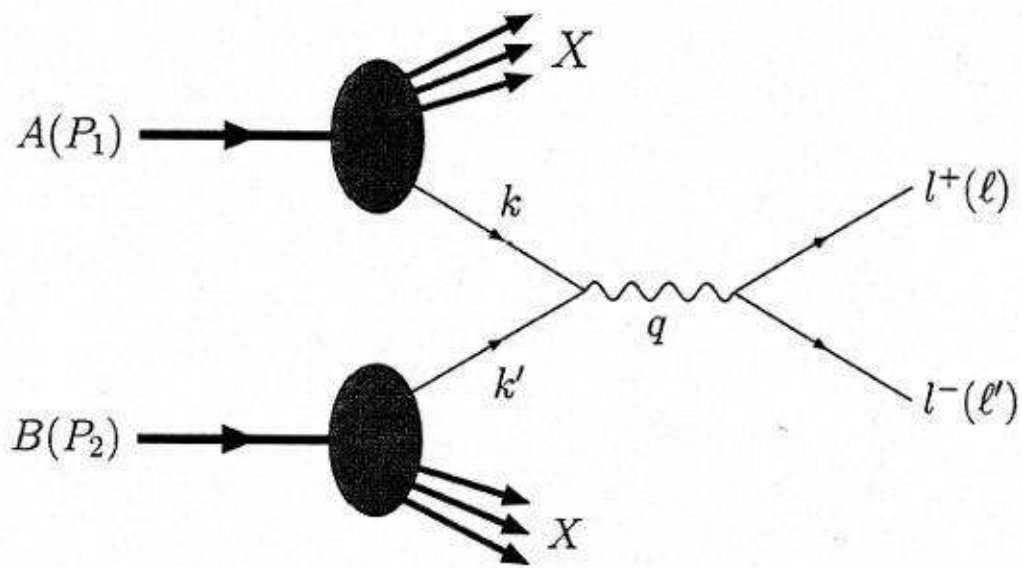}
\caption{Drell-Yan dilepton production.}

\label{fig:DrellYan}

\end{figure}

The antiproton probe is an ideal tool for such studies as it can be seen looking to two fundamental
Feynman diagrams describing 
the $\bar{p} p$ interaction. The first of them (Figure~\ref{fig:DrellYan}) is proportional 
to $\alpha^2$ and is 
the leading diagram illustrating the Drell-Yan process.

Here $k$ and $k^{'}$ are the quadrimomenta of the quark $q$ and antiquark $\bar{
q}$ annihilating into the virtual photon (with four-momentum q) that 
originates the lepton pair with the invariant mass $M$. Therefore at leading order the
 Drell-Yan cross section will depend only on the quark (antiquark) 
distributions in the colliding hadrons and on the cross-section of the elementary
 $q \bar{q} \rightarrow \gamma^{*}$ process.

\begin{figure} 
\centering
\includegraphics[totalheight=8.5cm]{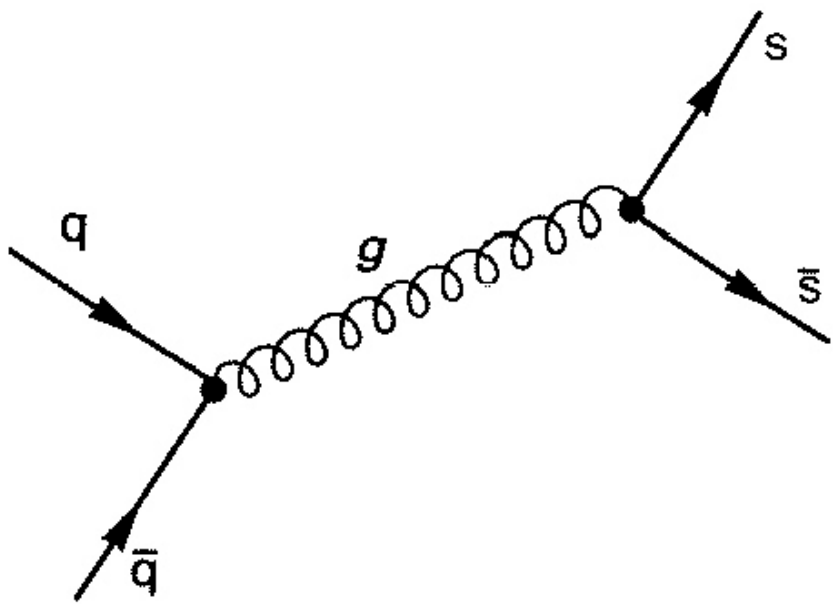}
\caption{Hyperons Production}

\label{fig:gssbar}

\end{figure}

The advantage of the Drell-Yan process versus other interactions like semi-inclusive deep inelastic scattering (SIDIS), 
for example, is then that in this process the spin dependent cross-sections can be expressed directly in terms of PD's 
and not as a convolution of PD's with quark fragmentation functions (QFF) like in the other hard processes. 
Therefore the interpretation of the data is simpler.
Compared to other probes (like pions, for example) the antiprotons are unique as each of its valence quarks 
can contribute to this diagram. 

Our purpose is then to exploit these advantages of the antiproton probe to measure PD's, in a complete range of the
Bjorken variable $x$, using an antiproton beam slowly extracted from SIS300 and a polarised (both
longitudinally and transversely) nucleon target. The relationship between the spin observables, that can so be measured
and the PD's is given in (Section~\ref{section:spinDY}).  

Of course a complete experiment would require also a
polarised antiproton beam and therefore we are deeply interested to contribute to all the initiatives, to
produce such beams either with the FILTEX or Stern-Gerlach method or making use of the $\bar{\Lambda} \rightarrow \bar{p} \pi^+$
decay, that provides polarised antiprotons. However, as it will be described later on, new and important results 
could be already obtained, namely for transverse distributions\cite{bar02}, with a polarised target and unpolarised antiproton beam, 
provided the energy requirements for this process will be fulfilled (Section~\ref{section:beamen}).
  
The second fundamental Feynman diagram, in the $\bar{p} p$ interaction that we refer to, is illustrated in 
(Figure~\ref{fig:gssbar}). Here too the antiproton probe is of particular interest as the correlation between the
$s$ and the $\bar{s}$ quarks can be observed. For example, in the reaction $\bar{p} p \rightarrow \bar{\Lambda} \Lambda$, 
assuming the spin orientation of the $\Lambda$ ($\bar{\Lambda}$) is given by that of the $s$ ($\bar{s}$) quarks,
the measurement, on the event by event basis, of the correlated $\Lambda$ ($\bar{\Lambda}$) polarisation will provide 
such correlation. The $\Lambda$ ($\bar{\Lambda}$) polarisation can be determined through the measurement of the 
angular distributions of $p$ ($\bar{p}$) from the weak decays $\Lambda \rightarrow p \pi^-$ 
($\bar{\Lambda} \rightarrow \bar{p} \pi^+$) in the $\Lambda$ ($\bar{\Lambda}$) rest frame, that carry such information.     
In addition to such exclusive processes, semi-inclusive reactions like $\bar{p} p \rightarrow \bar{\Lambda} \Lambda X$,
for example, are also accessible. From the measurement of spin dependent cross-sections of such reactions one
can extract either the QFF's or the PD's. This topic will be developed in (Section~\ref{section:spinhyp}).

Other important channels can be studied at the same time, measuring single spin asymmetries in semi-inclusive hadron
production like in the reaction $\bar{p} p \rightarrow \pi^+ \pi^- X$ as it will be discussed in (Section~\ref{section:singlespi}).

Last but not least spin dependent measurements will allow to disentangle the electric and magnetic part of the 
electromagnetic form factors in the $\bar{p} p \rightarrow l^+ l^- $ reaction (Section~\ref{section:formfac}).

\chapter{The Physics}
\section{Structure Functions}
\label{section:strucfunc}

At the twist two level $\boldsymbol{ \mathcal{O}(1)}$, the quark structure of the nucleon\cite{Jaf92} is described 
by three distribution functions. These can be also expressed in terms of the quark-nucleon forward scattering 
helicity amplitudes $A(h, H \rightarrow h^{'}, H^{'}$ where $h \, ( h^{'}) $ and $H \, ( H^{'})$ are the helicities 
of the quark and of the nucleon before (after) the interaction:

\begin{enumerate}

\item

the number density or unpolarised function, ${\it f_1(x)}$, that is the probability of finding a quark with a fraction 
{\it x} of the longitudinal momentum of the parent hadron, regardless of its spin orientation.

\begin{equation} \label{E:funo}
f_1(x) \; \propto \; A\left( \frac{1}{2} \; \; \frac{1}{2} \; \rightarrow \; \frac{1}{2} \; \; \frac{1}{2} 
\right) + A\left( \frac{1}{2} \; \; - \frac{1}{2} \; \rightarrow \; \frac{1}{2} \; \; -\frac{1}{2} \right)
\end{equation}
\item

the longitudinal polarisation, or helicity, distribution $ {\it g_1(x)}$, that measures the net helicity of a quark in a 
longitudinally polarised hadron, that is, the number density of quarks with momentum fraction {\it x} and spin parallel to 
that of the hadron minus the number density of quarks with the same {\it x} but spin antiparallel.

\begin{equation} \label{E:guno}
g_1(x) \; \propto \; A\left( \frac{1}{2} \; \; \frac{1}{2} \; \rightarrow \; \frac{1}{2} \; \; \frac{1}{2} 
\right) - A\left( \frac{1}{2} \; \; - \frac{1}{2} \; \rightarrow \; \frac{1}{2} \; \; -\frac{1}{2} \right)
\end{equation}
\item

the transverse polarisation distribution $ {\it h_1(x)}$, that, in a transversely polarised hadron, is the number density
of quarks with momentum fraction $x$ and polarisation parallel to that of the hadron minus the number density of quark with the same $x$ and antiparallel polarisation.

\begin{equation} \label{E:huno}
h_1(x) \; \propto \; A\left( - \frac{1}{2} \; \; \frac{1}{2} \; \rightarrow \; \frac{1}{2} \; \; - \frac{1}{2} 
\right) 
\end{equation}
\end{enumerate}

If quarks were perfectly collinear the distributions ${\it f_1(x)}$, $ {\it g_1(x)}$ and $ {\it h_1(x)}$ would exaust the
 information on the internal dynamics of the nucleon. If we admit instead a finite quark
transverse momentum $\boldsymbol{\kappa}_{\perp}$, the number of distribution functions increases. At twist two 
($\boldsymbol{ \mathcal{O}(1)}$) and three ($\boldsymbol{ \mathcal{O}(1/Q)}$) there 
are eight $\boldsymbol{\kappa}_{\perp}$ - dependent distributions. Three of them, upon integration over 
$\boldsymbol{\kappa}^{2}_{\perp}$,  
yield ${\it f_1(x)}$, $ {\it g_1(x)}$ and $ {\it h_1(x)}$. Other three, called 
$g_{1 T} (x,\boldsymbol{\kappa}^{2}_{\perp})$, $h^{\perp}_{1 L} (x,\boldsymbol{\kappa}^{2}_{\perp})$
and $h^{\perp}_{1 T} (x,\boldsymbol{\kappa}^{2}_{\perp})$, disappear when the hadronic tensor is integrated over 
$\boldsymbol{\kappa}_{\perp}$, as is the case in DIS. Finally two
T odd functions, named $f^{\perp}_{1 T}(x,\boldsymbol{\kappa}^{2}_{\perp}) $ and
$h^{\perp}_{1}(x,\boldsymbol{\kappa}^{2}_{\perp}) $, are respectively the distribution functions of an unpolarised quark 
inside a transversely polarised parent hadron and of a transversely polarised quark inside an unpolarised parent hadron.

To describe hadron production processes, other dynamical quantities are needed: the fragmentation functions, that describe
the probability for a quark, in a given polarisation state, to fragment into a hadron carrying some momentum fraction $z$. 
For them we will take the same notation as for the structure functions but using capital letters.

\section{Beam Energy for Drell-Yan processes}
\label{section:beamen}

The cross-section of the Drell-Yan reaction $\bar{p} p \rightarrow \mu^+ \mu^- X$, 
for a dimuon mass $M$, is given by:
 
\[
\frac{d^2 \sigma}{d M^2 dx_F} \; = \; \frac{4 \pi \alpha^2}{9 M^2 s} \frac{1}{(x
_1 \; + \; x_2)} \sum_{a} e^{2}_a \, [f^{a}(x_1)
\bar{f}^a (x_2) \; + \;\bar{f}^a (x_1) f^a (x_2) ]
\]
\[
\mbox{Here \quad} x_1 \; = \; \frac{M^2}{2 P_1 \cdot q} \mbox{\quad ; \quad} x_2
 \; = \; \frac{M^2}{2 P_2 \cdot q} \mbox{\quad that, in the parton model, are fractions of the }
\]

longitudinal momenta of the hadrons $A$ and $B$ carried by the quark and antiquark that annihilate into the virtual
photon. $ x_F \; = \; (x_1 \; - \; x_2)$ is 
the ratio of the longitudinal momentum of the pair to the maximum allowable momentum in the center of mass frame, and $a$ is the quark flavour, whereas

\begin{flushleft}
\[
 \tau \; = \; x_1 x_2 \; = \; \frac {M^2}{s}.
\]

\end{flushleft}

The scaling properties and the kinematical behaviour of the $ \bar{p} p \rightarrow \mu^+ \mu^- X$ reaction
are the same as for the $p p \rightarrow \mu^+ \mu^- X$, that has been successfully 
used to investigate the nucleon structure, mainly studying the $\bar{d} / \bar{u}$ asymmetry 
in the nucleon sea and testing the validity of the Gottfried integral.

\begin{figure}[h]
\centering
\includegraphics[totalheight=12cm]{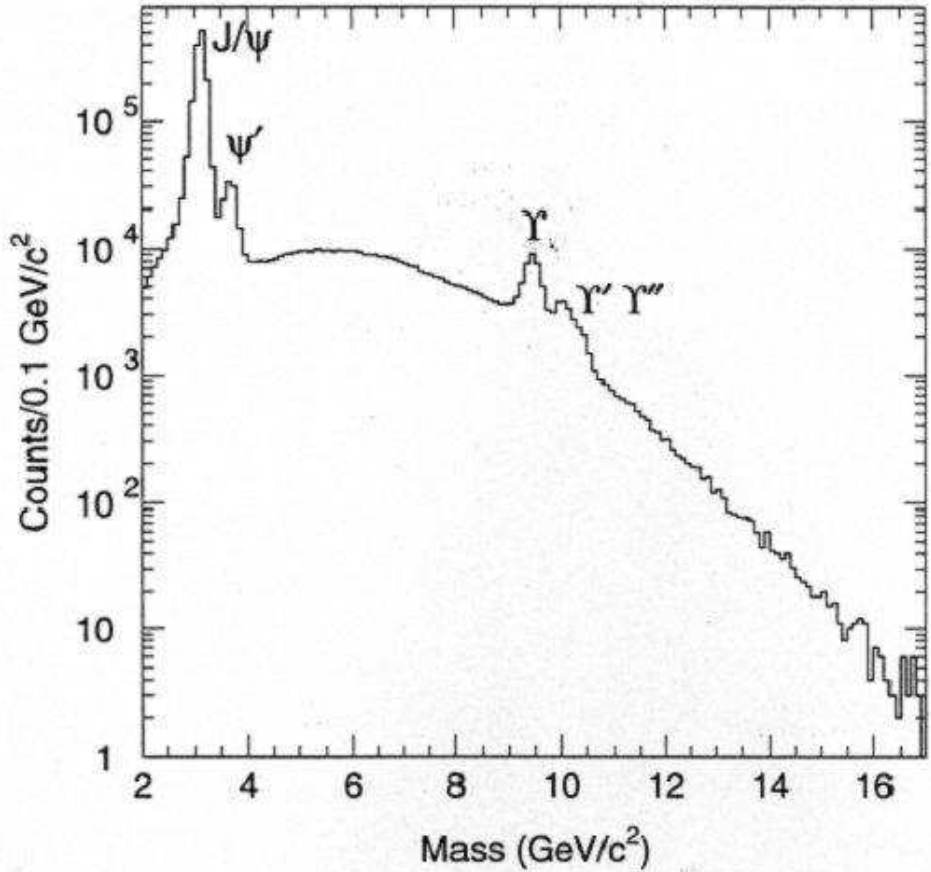}
\caption{Combined dimuon mass spectrum from $pp$ and $pd$ collisions (from\cite{E866}).}

\label{fig:future:Dimuon}

\end{figure}

The DY cross-section for the $pp \rightarrow \mu^+ \mu^- X$ reaction scales as:
$d^2 \sigma / d \sqrt{\tau} dx_F \, \varpropto \, 1 / s$. 
This greatly favours the lowest beam energy consistent with the selection of a dimuon pair mass,
(a dimuon mass spectrum \cite{Moss,E866} is shown in Figure~\ref{fig:future:Dimuon}),
produced in the ``safe'' region. 
The dimuon mass region, 
corresponding to values of $M$ ranging from 4 to 9 $GeV/c^2$, that is in between the $ J / \Psi$ and $\Upsilon$ 
resonance families, is called ``safe'' because there the dimuon spectrum
is essentially continuum and there are no resonance effects to disentangle in the interpretation of the data. 
For masses below the $ J / \Psi$ resonance families a number of potential backgrounds makes the understanding of 
the continuum more complex. 

The upper limit of 9 $GeV/c^2$ for $M$ defines the value of $s$ needed to get 
$0< \tau <1 $, that is to cover the full kinematical range allowed for $x_1$ and $x_2$. The lowest corresponding kinetic 
energy for the $\bar{p}$ beam is 40 GeV. This value is a reasonable compromise between the scaling behaviour of 
the cross-section and the need to cover the full range of the structure functions.

Scaling the data of
Ref. \cite{pbar125}, where the $ \bar{p}p \rightarrow \mu^+ \mu^- X$ reaction has been studied at the beam momentum 
of  125 GeV/c and for dimuon masses between 4 and 9 $ GeV / {c^2}$, we can expect an absolute cross-section, 
integrated over positive $x_F$ and all transverse momenta of about 0.3 nb at 40 GeV. This value will be used to estimate 
the expected counting rates. 

\section{Asymmetries in Drell-Yan processes}
\label{section:spinDY}

The Drell-Yan process, that finds out its best expression in the $\bar{p} p$ reaction, where all the valence quarks
can contribute, favours the study of chirally odd structure functions. This because in the standard parton
Drell-Yan diagram the two quarks chiralities are unrelated. Therefore there is not the chiral suppression of $h_1(x)$,
like in DIS. This reaction is then ideal to study transversity.

If both the beam and the target could be polarised either longitudinally or transversely the observable asymmetries and
their relations with the structure functions would be given by:

\begin{equation} \label{E:A_{LL}}
A_{LL} \; = \; \frac{\sum_a e^{2}_{a} g_{1}^a (x_1) g_{1}^{\bar{a}}(x_2)}{\sum_a e^{2}_{a} f_{1}^a (x_1) f_{1}^{\bar{a}}(x_2)}
\mbox{\quad for beam and target both longitudinally polarised \quad} 
\end{equation}
 
for beam and target both transversely polarised:

\begin{equation} \label{E:A_{TT}}
A_{TT} \; = \; \frac{sin^2 \, \theta \;cos \, 2 \phi}{1 \, + \, cos^2 \theta} \; 
\frac{\sum_a e^{2}_{a} h_{1}^a (x_1) h_{1}^{\bar{a}}(x_2)}{\sum_a e^{2}_{a} f_{1}^a (x_1) f_{1}^{\bar{a}}(x_2)}
\end{equation} 

and if one is longitudinal and the other transverse:

\begin{equation} \label{E:A_{LT}}
A_{LT} \; = \; \frac{2 \, sin \, 2 \theta \;cos \, \phi}{1 \, + \, cos^2 \theta} \; \frac{M}{\sqrt{Q^2}} \; 
\frac{\sum_a e^{2}_{a} (g_{1}^a (x_1) x_2 g_{T}^{\bar{a}}(x_2) \; - \; x_1 h_{L}^a (x_1) h_{1}^{\bar{a}} (x_2))}
{\sum_a e^{2}_{a} f_{1}^a (x_1) f_{1}^{\bar{a}}(x_2)}
\end{equation} 

\begin{figure}
\centering
\includegraphics[totalheight=7cm]{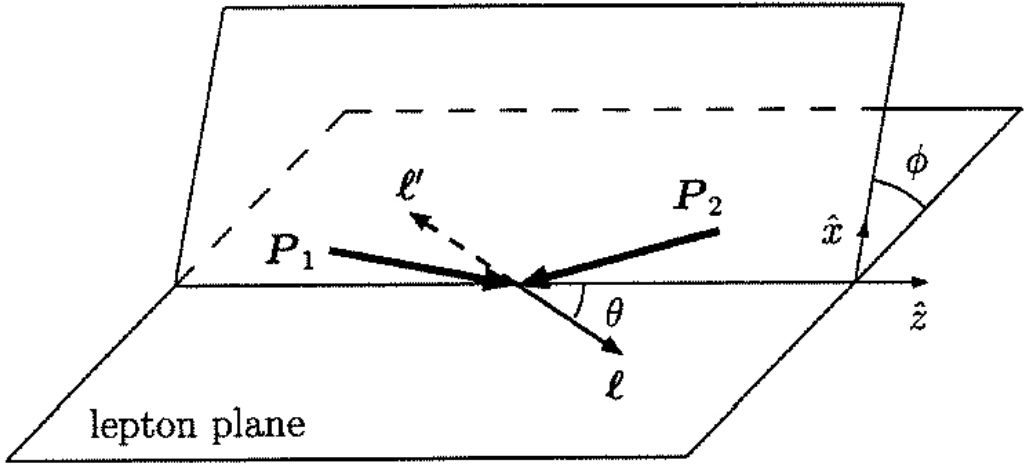}

\caption{The geometry of the Drell-Yan production in the rest frame of the lepton pair.}

\label{fig:future:Dilepto}

\end{figure}

Here the angles $\theta$ and $\phi$ are the polar and azimuthal angles defined in Figure~\ref{fig:future:Dilepto}.
To extract the structure functions these theoretical expression have to be compared in a fitting
procedure to the experimental asymmetries measured at the same values of $x_1(x_2)$. 
These asymmetries have to be corrected
by a factor
$1/(P_b \cdot f \cdot P_T)$ taking into account for the beam polarisation $P_b$, the dilution factor $f$ and the
target polarisation $P_T$. For an $NH_3$ polarised target, the dilution factor, that is the number of polarised
nucleons over the total number of nucleons in the target, $f \, = \, 0.176$ and $P_T \, = \, 0.85$.

\begin{figure}[h]
\centering
\includegraphics[totalheight=12cm]{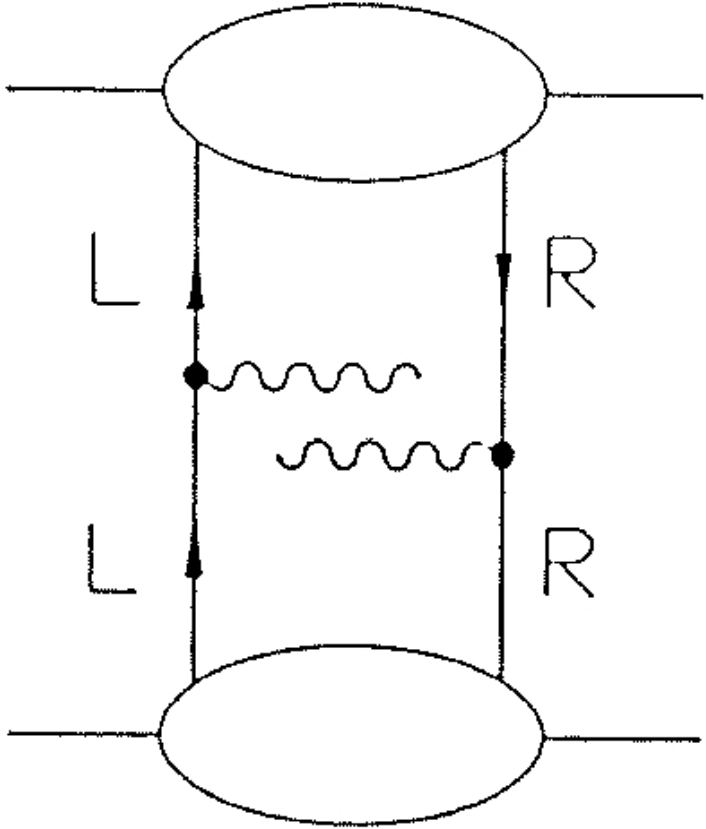}
\caption{Chirality in Drell-Yan production of lepton pairs.}

\label{fig:future:Chiodd}

\end{figure}

If both a polarised beam and a polarised target transversely polarised were available the Drell-Yan
reaction $\bar{p} p \rightarrow \mu^+ \mu^- X$ would be the ideal tool to study transversity, that is very
poorly known. This because $h_1^a (x)$ (see equation~\ref{E:A_{TT}}) has not to be unfolded 
with fragmentation functions, but also mainly because being a chirally odd function is not suppressed like in
DIS processes. This last property is illustrated in Figure~\ref{fig:future:Chiodd} (from Ref.~\cite{Jaf92})

However as the $\bar{p}$ polarised beam is not yet available we have looked to alternative ways to get
transverse distributions of quarks inside the nucleon.
 
The angular distribution of dileptons, for unpolarised beam and target is:

\begin{equation} \label{E:dsig}
\frac{1}{\sigma} \frac{d \sigma}{d \Omega} \; = \; \frac{3}{4 \pi} \frac{1}{\lambda \; + \; 3} \, \times
\, (1 \; + \; \lambda \cos^2
~\theta \; + \; \mu \sin^2 \theta \cos \phi \; + \; \frac{\nu}{2} \sin^2 \theta \cos 2 \phi )
\end{equation}

where $\theta$ is the polar angle of the lepton in the virtual photon rest frame as defined before
(see Figure~\ref{fig:future:Dilepto}).

Perturbative QCD calculations at next-to leading order give $\lambda \approx 1$, $\mu \approx 0$ , $ \nu
\approx 0$,
confirming the characteristic $\cos^2 \theta$ distribution of the decay of a transversely polarised 
virtual photon,
given in the parton model. However fits of experimental data (see, for example Ref.~\cite{Con89} show 
remarkably large values of $\nu$, reaching values of
about $30 \%$. Recently \cite{brod03,col03}
it has been pointed out that initial state interaction in the unpolarised Drell-Yan process could explain
the observed asymmmetries and be connected with the quark (antiquark) T-odd distributions 
$h^{\perp}_{1}(x_2 ,\boldsymbol{\kappa}^{2}_{\perp}) $ and 
$\bar{h}^{\perp}_{1}(x_1 ,\boldsymbol{\kappa}^{2}_{\perp}) $. Under this hypothesis, numerical estimations 
of the asymmetry for the $ \bar{p}p \rightarrow \mu^+ \mu^- X$ process, 
for the maximum of $\nu$, of $30 \%$ are given.

The measurement of $\cos 2 \phi$ contribution to the angular distributions of the dimuon pair of the 
$ \bar{p}p \rightarrow \mu^+ \mu^- X$ process will provide the product 
$h^{\perp}_{1}(x_2 ,\boldsymbol{\kappa}^{2}_{\perp}) \, \bar{h}^{\perp}_{1}(x_1,\boldsymbol{\kappa}^{' 2}_{\perp}) $.
We have already proposed this measurement for the PANDA detector, where a polarised target cannot be installed because
of the disturbance of the magnetic field of the solenoid. However the maximal antiproton beam energy available there
limits considerably the domain of Bjorken x reachable (see Section~\ref{section:beamen}). 

If a transversely polarised hydrogen target were available, the measured asymmetry for the two target spin states
depends on the $sin (\phi \; + \; \phi_{S_1})$ term, where $\phi_{S_1}$ is the azimuthal angle of the target spin in the
frame of Fig.~\ref{fig:future:Dilepto}. This term is
$\propto h_1(x_2 ,\boldsymbol{\kappa}_{\perp}^2 ) \, \bar{h}_{1}^{\perp} (x_1 , \boldsymbol{\kappa}_{\perp
}^{' 2})$, as shown by\cite{bo99}:

\begin{equation} \label{E:A_{T}}
A_{T} \; = \; \mid \boldsymbol{S_{\perp}} \mid\frac{2 \, sin \, 2 \theta \; sin \,( \phi \, - \, \phi_{S_1})}{1 \, + \, cos^2 \theta} \; \frac{M}{\sqrt{Q^2}} \; 
\frac{\sum_a e^{2}_{a}[x_1 (f_{1}^{a \perp}(x_1)  f_{1}^{\bar{a}}(x_2) \; + \; x_2 h_{1}^a (x_1) h_{1}^{\bar{a} \perp} (x_2))]}
{\sum_a e^{2}_{a} f_{1}^a (x_1) f_{1}^{\bar{a}}(x_2)}
\end{equation} 

This measurement, in the absence of a polarised beam, is a unique tool to
probe the $\boldsymbol{\kappa}_{\perp}$ dependence of quark distributions inside the nucleon.

\section{Spin Asymmetries in Hyperon Production}

\label{section:spinhyp}

 In 1976 it was discovered that inclusively produced \lam's, in unpolarized
$p p$ interactions, are negatively polarized in the direction
normal to production plane. The magnitude of the polarization rises
with $x_F$ and $p_T$ and achieves ~40\%. Even higher \lam's polarization
(~60\%) was obtained in exclusive reactions like $pp
\rightarrow p \Lambda K^+$, $pp \rightarrow p \Lambda K^+ \pi^+
\pi^-$ {\it etc}~\cite{R608,felix}. Since then this phenomenon was
confirmed many times in extensive set of experiments but its
theoretical explanation still remains a persisting problem.

Different quark-parton models using static SU(6) wave functions
were proposed to interpret these polarization effects by
introducing a spin dependence into the partonic fragmentation and
recombination processes~\cite{degr,string,szwed}. The
\lam~polarization is attributed to some mechanism, based on
semiclassical arguments~\cite{degr,string} or inspired by
QCD~\cite{szwed}, by which produced strange quarks acquire a large
negative polarization. Recently a new approach to this problem
based on perturbative QCD and its factorization theorems, and
which includes spin and transverse momentum of hadrons in the
quark fragmentation, was proposed in~\cite{anselm}. These models are
based on different assumptions and are able to explain the main
features of the \lam~polarization in unpolarized $pp$-collisions. To
better distinguish between these models more complex phenomena have
to be considered.

With polarized beams or target one can access new supplementary
observables -- the analyzing power, $A_N$, and the depolarization
(sometime referred as spin transfer coefficient) , $D_{NN}$,
defined as

\begin{equation}
\label{eq:an}
A_N = \frac{1}{P_B cos\phi} \frac{N_{\uparrow}(\phi)
- N_{\downarrow}(\phi)}{N_{\uparrow}(\phi) +
N_{\downarrow}(\phi)},
\end{equation}

\begin{equation}
\label{eq:dnn} D_{NN} = \frac{1}{2P_B cos\phi}
[P_{\Lambda\uparrow}(1+P_B A_N cos(\phi)) -
P_{\Lambda\downarrow}(1-P_B A_N cos(\phi))],
\end{equation}
where $\phi$ is the azimuthal angle between the beam polarization
direction and the normal to scattering plane.

It is interesting to note that whereas produced \lam~polarization
remains large and negative for exclusive and inclusive channels
the spin transfer coefficient is negative in low energy (beam
momentum 3.67 GeV/c) exclusive production ~\cite{disto},
compatible with zero at intermediate energies (beam momentum 13.3
and 18.5 GeV/c)~\cite{ags} and positive at high energy (beam
momentum 200 GeV/c) inclusive reaction~\cite{bravar}. Thus, the
measurements  at 40 GeV/c can bring an additional information on
this phenomenon.

The spin dependence of exclusive annihilation reaction $\bar p +p
\rightarrow \bar \Lambda+\Lambda$ has been considered relevant
to the problem of the intrinsic strangeness component of
nucleon~\cite{alberg}-\cite{rekalo}. It was
demonstrated~\cite{paschke} that the use of a transversely
polarized target, in principle, allows the complete determination
of the spin structure of the reaction. Corresponding measurements
was performed by PS185 Collaboration, see~\cite{ps185} and
references therein. Competing models such as $t$-channel meson
exchange model and $s$-channel constituent quark model reasonably
well describing the cross-section of this reaction exist. But both are
unable to describe such spin observables as spin transfer from
polarized proton to \lam~ ($D_{NN}$) and to \alam~ ($K_{NN}$). It
is evident that new data on spin transfer and correlation
coefficients at higher energies and momentum transfer will be
easier interpret in QCD based approaches and can help us to better
understand the spin dynamics of strong interactions.

\subsection{Advantages of the antiproton probe}

One of the advantages of the Drell-Yan process versus other hard processes
was discussed in section~\ref{section:spinDY}: in the standard parton
Drell-Yan diagram the two quarks chiralities are unrelated (see Figure~\ref{fig:future:Chiodd}). 
Therefore there is not the chiral 
suppression of $h_1(x)$,like in DIS.
A similar situation exists in the process
$\bar{p} p \rightarrow \Lambda \bar{\Lambda} X$, described by the diagram of 
Figure~\ref{fig:future:Chiodd}. Here the basic diagram, illustrating the chirality conservation,
is that of Figure~\ref{fig:gssbar} where a gluon replaces the $\gamma$. Therefore here
too the two quarks chiralities are unrelated and there is not a chiral suppression
of $h_1(x)$, like in DIS.
Another advantage of the use of the antiproton beam is related 
to a complete correlation between the spin orientations of the $\Lambda$($\bar{\Lambda}$).

In addition the $\Lambda$($\bar{\Lambda}$) polarisations can be easily extracted from the
angular distributions of the weak decays of the $\Lambda \rightarrow p \pi^-$
($\bar{\Lambda} \rightarrow \bar{p} \pi^+$). These decays have both a large asymmetry parameter
($\alpha \, = \, 0.642$) and branching ratio ($B. \, R. \, = \, 0.640$). Therefore, even with an
unpolarised antiproton beam but with a polarised target one can get the spin
correlation parameters related both to the PD's and to the QFF's.

\section{Single Spin Asymmetries}

\label{section:singlespi}

The perturbative QCD spin dynamics, with the helicity conserving
quark-gluon couplings, is very simple. However, such a simplicity
does not appear in the hadronic spin observables; the QCD spin
structure is much richer and more surprising than one could naively
expect from the underlying parton dynamics. The non perturbative,
long distance QCD physics has many spin properties, yet to be explored:
the transverse quark spin distribution is unknown; subtle spin effects
related to parton intrinsic motions in distribution functions and in
fragmentation processes have been proposed and might be responsible
for observed spin asymmetries; a QCD spin phenomenology seems to be
possible, but more data and new measurements are crucially needed.

A typical example of such aspect of QCD is supplied by the transverse
Single Spin Asymmetries (SSA),
$$
A_N = \frac{d\sigma^\uparrow - d\sigma^\downarrow}
{d\sigma^\uparrow + d\sigma^\downarrow} \>,
$$
measured in $p^{\uparrow} \, p \to \pi \, X$
and $\bar p^{\uparrow} \, p \to \pi \, X$ processes: the SSA at large
values of $x_F$ ($x_F \gtrsim 0.4$) and moderate values of $p_T$
($0.7 < p_T < 2.0$ GeV/$c$) have been found by several experiments
\cite{ags,e704,star} to be unexpectedly large. These asymmetries have
clear features:
\begin{itemize}
\item
the pion production at large $x_F$ values originates from valence
quarks, and indeed the sign of $A_N$ (positive for $\pi^+$ and negative
for $\pi^-$) reflects the expected sign of $u$ and $d$ quark polarization;
\item
similar values and trends of $A_N$ have been found in experiments with
center of mass energies ranging from 6.6 up to 200 GeV: this seems to hint
at an origin of $A_N$ related to fundamental properties of quark distribution
and/or fragmentation.
\end{itemize}

A new experiment with anti-protons scattered off a polarized proton target,
in a new kinematical region, could certainly add information on such spin
properties of QCD. Also, $A_N$ observed in
$\bar p \, p^{\uparrow} \to \pi \, X$ processes should be related to $A_N$
observed in $\bar p^{\uparrow} \, p \to \pi \, X$ reactions, which should be
checked.

Recently, several papers have stressed the importance of measuring SSA in
Drell-Yan processes \cite{hts}-\cite{adm}; these measurements allow the
determination of new non perturbative spin properties of the proton, which 
gives the distribution of quarks in a transversely
polarized proton \cite{adm}. These measurements can, at the moment, only be
performed at RHIC, in the collision of transversely polarized protons on
unpolarized protons; however, as it was discussed in Chapter 2, Drell-Yan
processes in $pp$ interactions, at large values of $s$, are much less abundant
than in $\bar pp$ interactions, at smaller values of $s$. In that the study of
$\bar p^{\uparrow} \, p \to \mu^+ \, \mu^- \, X$  processes at GSI offers
unique possibilities.

\section{Electromagnetic Form Factors}
\label{section:formfac}
Nucleon electromagnetic form factors (FFs) in the
time-like (TL) region, which can be measured through the reactions $\overline{p}
+p\leftrightarrow e^+ +e^-$, contain important information about the nucleon 
structure, which can not be obtained from elastic $eN$-scattering, i.e in  
space-like (SL) region. First of all these FFs are complex functions of $s=-q^2$, in TL
region, and analyticity allows to connect the two regions  through dispersion 
relations.

The data about FFs in TL region, although they extend at higher $s$, are less 
precise in comparison with SL region: statistics is very low, measurements of 
angular distribution are very scarce and unprecise (absent for neutron), and experimental 
information on  polarization phenomena is absent.

The reaction $\bar{p} p \rightarrow \mu^+ \mu^-$ can be an alternative way to study
FFs measuring both the angular distributions of the differential cross-sections and of
the analysing power.

One can express the angular dependence of the differential cross section for 
$\overline{p}+p\to \mu^+ +\mu^-$ as a
function of the angular asymmetry ${\cal R}$ as:
\begin{equation}
\displaystyle\frac{d\sigma}{d(\cos\theta)}=
\sigma_0\left [ 1+{\cal R} \cos^2\theta \right ],~
{\cal R}=\displaystyle\frac{\tau|G_M|^2-|G_E|^2}{\tau|G_M|^2+|G_E|^2}
\label{eq:eq3}
\end{equation}
where $\sigma_0$ is the value of the differential cross section at
$\theta=\pi/2$.

These quantities are very sensitive to the different underlying
assumptions about the $s$-dependence of the FFs, therefore a precise measurement
would be very interesting.

The measurement of the differential
cross section for the process $\overline{p}+p\to e^+ +e^-$ at a fixed value of
$s$
and for two different angles $\theta$,  allowing  the separation of the two FFs,
 $|G_M|^2$ and $|G_E|^2$, is equivalent to the well known Rosenbluth separation
for
the elastic $ep$-scattering. However in TL, this procedure is simpler, as it
requires to change only one kinematical variable, $\cos\theta$, whereas, in SL
it is
necessary to change simultaneously two kinematical variables: the energy of the
initial electron and the electron scattering angle, fixing the momentum transfer
square, $q^2$.

The angular dependence of the cross section, Eq. \ref{eq:eq3}, results
directly from the assumption of one-photon exchange, where the spin of the
photon
is equal 1 and the electromagnetic hadron interaction satisfies the
$C-$invariance.
Therefore the measurement of the differential
cross section at three angles (or more) would also allow to test the presence of
$2\gamma$ exchange.

Polarization phenomena will be especially interesting in $\bar{p} p \rightarrow \mu^+ \mu^-$. 
For example, the transverse polarization $P_T$ of the proton target 
(or transverse polarization of the antiproton beam) results in nonzero analyzing power:
$${\cal A}=\displaystyle\frac{\sin 2\theta Im G_E^*G_M}{D\sqrt{\tau}},
~D=|G_M|^2(1+\cos^2\theta)+\displaystyle\frac{1}{\tau}|G_E|^2\sin^2\theta$$
$$
\displaystyle\frac{d\sigma}{d\Omega}(P_T)=
\left ( \displaystyle\frac{d\sigma}{d\Omega} \right )_0 \left [1+{\cal A}P_T
\right ]
$$
This analyzing power characterizes the T-odd correlation 
$\vec P\cdot\vec k\times\vec p$, where $\vec k(\vec p)$ is the three momentum 
of the $\overline{p}$ beam (produced lepton). It is important to note that the 
$\tau$-dependence of ${\cal A}$ is very sensitive to existing models of the nucleon FFs, 
which reproduce equally well the data in SL region.

The same information can be obtained from the final polarization in 
$e^++e^- \to \vec p+\overline{p}$, but in this case one has to deal with the problem of 
hadron polarimetry, in conditions of very small cross sections.

The main problems that can be solved by future measurements with a polarized ant
iproton beam (or an unpolarized proton beam on a polarized proton target), in view 
of a global interpretation of the four nucleon FFs (electric and magnetic, 
for neutron and proton) in TL and SL momentum transfer regions, are:

\begin{itemize}

\item The separation of the electric and magnetic FFs, through the angular 
distribution of the produced leptons: the measurement of the asymmetry ${\cal R}$ 
(from the angular dependence of the
differential cross section for $\overline{p}+p\leftrightarrow e^+ +e^-$) is
sensitive to the relative value of $G_M$ and $G_E$.

\item The presence of a large relative phase of magnetic and electric proton FF
in the TL region, if experimentally proved at relatively large momentum transfer,
can be considered a strong
indication that these FFs have a different behavior.

\end{itemize}

\chapter{Antiproton production and experimental set-up }
\label{section:setup}
\section{Beam and target}
\label{section:beamtarget}

The announced performances of HESR give a reference antiproton beam energy of 14.5 GeV, a
luminosity $\le 2 \times 10^{32} \; cm^{-2} s^{-1}$ and a momentum spread lower than $\pm 1 \times 10^{-4}$.
These excellent performances do not fit with the experimental program we propose, that requires a minimal
energy of 40 GeV and a limited momentum resolution. The present design of the PANDA detector, in addition,
excludes the possibility to replace the internal unpolarised target with a polarised one.

We propose then to slowly extract an antiproton beam of $\geq40$ GeV/c from SIS 300 to hit an external polarised target.
The expected momentum spread of such a beam should be about $\pm 2 \times 10^{-4}$, that is largely enough for
our experiment.

To estimate the number of antiprotons that we can get from such extraction we assume the whole foreseen antiproton production
accumulation rate of $7 \times 10^{10} \; \bar{p} / h$ to be available and that the injection and extraction efficiencies
will be always larger than $0.90$. Within these assumption, we can expect a beam intensity on the target of about
$1.5 \times 10^{7} \; \bar{p} \cdot s^{-1}$.

For the polarised target we take as an example the polarised target system now in use in COMPASS at CERN where two
cells with opposite polarisation are put one downstream of the other. This solution allows to minimize the systematic errors.

The expected luminosity will be, for a $NH_3$ target $10 \; g/cm^2$ thick:

$$\boldsymbol{ \mathcal{L}} \; = \; \frac{3}{17} \times 10 \times 6 \cdot 10 ^{23} \times 1.5 \cdot 10^7 \; = \; 1.5 \cdot 10^{31} \; \, cm^{-2} \, s^{-1}$$

For completeness we recall that the dilution factor for such a target is $f \, = \, 0.176$ and thr polarisation
$P_T \, = \, 0.85$.

The generation of a 40 GeV/c antiproton beam would require the following
additional construction or modification items to the presently proposed configuration scheme of the
new International Accelerator Facility in GSI:
\begin{enumerate}
\item
Extraction of the accelerated anti proton beam from SIS 100 into SIS 300.
Such a transition system would need to be designed and built. Or
alternatively an injection scheme from the CR into the SIS 300.
\item
A slow extraction system from SIS 300 into a more powerful extraction
beamline able to handle momenta larger than 40 GeV/c.
\item
a cave housing the experimental setup as proposed which
can handle the expected radiation doses (with $\sim 2 \times 10 ^7 \; \bar{p} \cdot s^{-1}$).
\end{enumerate}

With this scheme, and provided that different spin orientations were available for the polarised target like in the
COMPASS set-up, both longitudinal and transverse asymmetries could be measured. In addition if a transversely
polarised antiproton beam could be produced and extracted from SIS 300 a unique tool for the study of the nucleon 
structure would be available.

An alternative solution, proposed by Hans Gutbrod, could be to imagine the HESR as a collider with both a 
polarised proton  and antiproton beams interacting with a luminosity comparable to that reachable with an 
external target.
If such a luminosity could be reached the advantage of such a solution would be that  no dilution 
factor has to be considered in this case making the asymmetries measurement more precise for the same number 
of events collected. However to get a proton polarisation $P_p \, = \, 0.85$ would be a
difficult task to achieve. In addition only transverse asymmetries could be measured this way. 
The required CM energy $\sqrt{s}$ for the proposed program could be easily reached with the present foreseen 
performances of HESR (15 GeV/c).
In addition the higher CM energies available would allow new physics opportunities.

A polarized proton beam of up to 15 GeV/c would require a polarized proton
source and an acceleration scheme preserving the polarization. No new beam
line needs to be built and no additional extraction needs to be included
into the acceleration system. The lattice of the HESR would have to allow an
interaction region of both beams.

One of the key issues of all these proposals is the luminsity.
We are aware of efforts to improve the overall antiproton production rate. We strongly support these efforts.

If the collider solution should be choosen the design of the experimental set-up would be
then entirelly different.
%

%
%\subsection{Expected Counting Rates}
%
%The total available beam for experiments is limited by the production yield of 1.5 $\times$10$^7$/s 
%antiprotons. 
\section{Instrumentation and Detectors}

\begin{figure}
\centering
\includegraphics[totalheight=9.5cm]{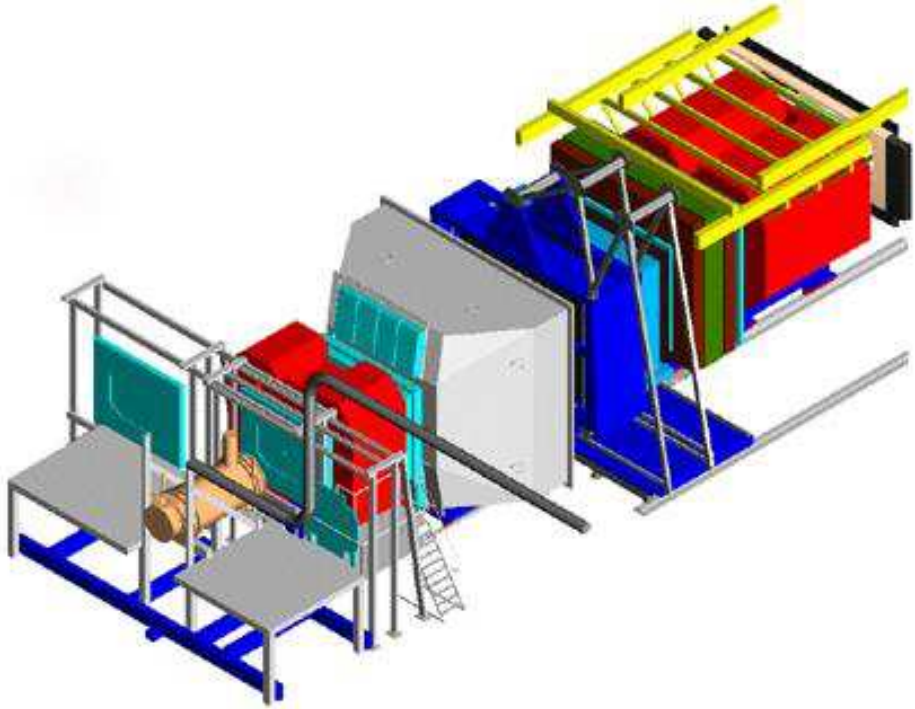}
\caption{The first part of the COMPASS spectrometer.}

\label{fig:Compass}

\end{figure}

The proposed detector concept is inspired from the Large Angle Spectrometer, that is the first part of the COMPASS
spectrometer, shown in Figure~\ref{fig:Compass} .This choice is just made to show the
feasibility of the experiment with an apparatus that we know well, as a large part of us is using it now. Of course if the collider mode should become available, a different set-up should be foreseen. 

\section{Overview of the detector concept}
\label{section:detector:concept}
The Large Angle Spectrometer consists of a large dipole magnet SM1 and various kinds of tracking 
detectors. 
SM1 is a window frame magnet with an aperture of 2.0 x 1.6 for a depth of about 1 m. It
provides a field integral of about 1 Tesla m.

The tracking detectors have been chosen in such a way that they can sustain the beam rate 
($1.5 \times 10^7 \; \bar{p} / s.$) and provide the hits position with such a precision to guarantee the needed
resolution for the position of the vertices of the decaying particles ($\Lambda$ and $\bar{\Lambda}$) 
and for the widths of the corresponding peaks in the invariant mass spectra.
To reach these goals and also to minimize the overall cost of the apparatus, detectors of smaller size
but with thinner resolution and accepting higher rates have been choosen to detect the hits
nearer to the beam trajectory. These detectors are GEM and MICROMEGAS, that provide spatial resolutions
with $\sigma \leq 70 \,\mu $. To detect hits at larger distances from beam trajectories MWPC and STRAW tubes
are used that provide spatial resolutions $> 1.5 $ mm. These last detectors have a dead zone in their central
part, that nearer to the beam trajectories and covered by the GEM and MICROMEGAS detectors.

With this setup a mass resolution ($\sigma \approx 2.5 \; MeV/c^2$) can be obtained for the $\Lambda$ 
($\bar{\Lambda}$).

The expected spatial resolution on the position of the decay vertices of the $\Lambda$ ($\bar{\Lambda}$)
goes from $\approx 1 \; cm$., for very small angles with the beam trajectories, to a couple of mm. for larger
angles. This spatial resolution is large enough to base the $\Lambda$ ($\bar{\Lambda}$) identification
on the requirement that these vertices are outside the target.

An hodoscope of scintillators will provide the trigger with the requirement that at least two hits are present.

For the muon detection sandwiches of iron plates, IAROCCI tubes and scintillator slabs are foreseen. 
The scintillating slabs will provide the trigger with the requirement that two hits are present.

Finally, to minimize the background related to the interactions of the beam after the target, a vacuum
pipe of growing cross-section, will catch the beam up to the beam dump. This solution, that takes advantage
of the good emittance of the extracted beam, was already successfully applied in the DISTO experiment.
\section{Summary and Conclusion}

We hope to have underlined what a rich information on the nucleon structure could be obtained from the
measurements we propose, the ideal tools being both a polarised antiproton beam and a polarised nucleon target.

As already mentioned, to get structure functions in a large x Bjorken domain the total energy in the C.M. system
$\sqrt s$ is a key issue. This can be obtained in two ways: either with a slow extraction from SIS300 of an antiproton
beam with momenta $\ge \, 40 GeV/c$ (polarised?) or, as proposed by H. Gutbrod, using HESR as a collider of antiprotons
(polarised?) and transversely polarised protons ( we skip here the possibility to have longitudinal polarisation ).

In the first case with a polarised target having both longitudinal and transverse polarisation available, like
the COMPASS target for example, one can collect more information as discussed in Section~\ref{section:spinDY}.
This will require from GSI the additional modifications discussed in Section~\ref{section:beamtarget}.

The collider solution will have the advantage, for an equal luminosity, of a better factor of merit, for
a proton polarisation equal to that of the target, as no dilution factor has to be taken into account in that case.
The modification required by GSI are also indicated in Section~\ref{section:beamtarget}.

From the detector point of view the collider solution is slightly better as a larger acceptance can be obtained.
this because, the cut in acceptance for very forward (backward) emitted pairs ( of muons or Lambda's ) could be
smaller.

To go further in this project, the collaboration needs to know from GSI what is feasible and discuss with
the management the possible issues, once the physical case has been evaluated.  

\end{document}